# User Experience in Human-Machine Interaction: Insights from Field Studies in Autonomous Mobility

Helen Schneider[1], Svetlana Pavlitska[1,2], and J. Marius Zöllner[1,2]

[1] Karlsruhe Institute of Technology (KIT), Germany
[2] FZI Research Center for Information Technology, Germany
helen.schneider@kit.edu

**Abstract.** Autonomous vehicles (AVs) promise safer, cleaner, and more inclusive mobility, yet large-scale adoption is hindered by user acceptance rather than by technical challenges. Prior studies on acceptance and user experience largely rely on surveys, simulators, or Wizard-of-Oz setups, often over-representing technologically enthusiastic participants and focusing on drivers instead of passengers. We address this gap with real-world field studies with AVs in real traffic, totaling 144 participants. Using multi-modal sensing, we evaluated EEG, heartbeat, breathing, camera, and voice signals for affect inference in combination with vehicle data. Our results show that breathing, camera, and voice measurements are reliable and practical in naturalistic passenger contexts. We further contribute a validated study protocol, a self-assessment app for real-time assessment during human–machine interaction, and a tailored questionnaire to capture participant attitudes towards AVs. By grounding UX evaluation in real-world contexts, this work lays a foundation for user-centered design of autonomous mobility systems and robotics in general. Our work bridges the gap between affective computing and technical implementation of autonomous vehicles.

## 1 INTRODUCTION

The ongoing advancements in the automotive industry have been significantly influenced in recent years by autonomous driving (AD) [22]. This innovation already plays a crucial role in reshaping how we commute and transform traffic dynamics within urban areas [7]. Benefits of autonomous vehicles (AVs) are seen in numerous fields, e.g. the reduction of traffic accidents and CO2 emissions, and improved mobility of underage, older, or impaired persons [45].

Nonetheless, the general public still has reservations about AVs [19]. A key obstacle to broad public acceptance is not only of a technical nature, but rather a psychological one [39]. While there are several works on acceptance factors of AVs and driver-data collection [10, 24, 34, 41], most studies are based on general surveys, focused on driver behavior instead of passengers, and not linked to field AD experiments [27].

Due to limited user acceptance, researchers are seeking ways to enhance the user experience (UX) during AV rides, as acceptance and UX are closely intertwined [26]. Automatic inference of UX has been done among others with facial configurations [43] and physiological signals (e.g. heartbeat [40], galvanic skin response [30], EEG [16]). To the best of our knowledge, no real world field-studies were conducted, which included passenger affect and autonomous driving data. There exist no study protocols to collect self-labeled affective states of users including multi-modal objective user-sensor-data. The aim of our work is therefore to *provide a real-world study protocol to gather multi-modal user sensor data and machine data collectively.* We conducted 8 real-world studies with 144 participants to evaluate the feasibility of electroencephalography (EEG), heartbeat, breathing, camera, and voice measurements for affect inference[3]. Our contribution is three-fold:

1. We identify breathing, camera, and voice measurements as feasible user data in real-world studies.
2. We provide a study setup for the research community to incorporate into their own development process in robotics, to generalize field study findings over the world population.
3. We provide a questionnaire to improve transparency into participants' attitudes toward AD and new technologies, as well as their UX during the experiments including the self developed SAMotion-app for real-time self-assessment during human-machine interaction (HMI).

## 2 RELATED WORK

### 2.1 Terminology

**User Experience (UX)** According to ISO Standard 9241-11 [18] UX is defined as *"[a] person's perceptions and responses that result from the use and / or anticipated use of a product, system or service."* Hassenzahl and Tractinsky [11] emphasize that UX goes far beyond merely fulfilling instrumental needs through technology. UX emerges from the user's inner state, the characteristics of the designed system, and the surrounding context or environment in which the interaction takes place. Consequently, UX research should not solely focus on avoiding encountered problems but should strive to create exceptional quality experiences. This challenges a central assumption of traditional HMI, where quality is equated with the absence of problems.

Our work focuses on assessing the *UX during direct HMI*, specifically the time spent in a moving AV. AV design features, comfort during boarding and exiting, and potential waiting times are not factored into participants' evaluations.

[3] The study protocol was approved by the Ethics Committee of Karlsruhe Institute of Technology (protocol code A2024-012, date of approval April 08, 2024. Informed consent was obtained from all subjects involved in the study.

**Affect for User Self-Assessment** There are three main approaches to modeling **affect**: discrete, continuous, and hybrid. Discrete models identify categories, e.g., six basic emotions such as anger, disgust, fear, happiness, sadness, and surprise [6]. Continuous models use continuous scales that mostly comprise at least the two dimensions of valence (negative/positive) and arousal (calm/excited), e.g., the circumplex model of affect [36], and sometimes incorporate a third dimension of dominance (no control/in control), e.g., the VAD model (valence-arousal-dominance) [37]. Hybrid models combine discrete and continuous affect modeling. We focus on VAD, based on the findings of Barrett et al. [1].

### 2.2 SoTA Datasets for affective computing and AD

There exist datasets to predict valence and arousal through modalities such as camera, audio, infrared, EEG, and additional physiological signals, such as galvanic skin response, heart rate variability, respiration rate, and blood pressure [38]. The most commonly used modalities in these datasets were camera and audio. However, these datasets do not include robotic data, limiting comprehensive evaluation and understanding of HMI. In the context of AD, there exist widely used datasets from simulation and real-world vehicle data [34]. However, they do not contain information on people's feelings towards AVs as either direct passengers or vulnerable road users.

Multi-modal datasets containing LiDAR, passenger-camera, biometrics, and self-assessment labels are scarce. Some works provide biometric driver data [34], but do not include passenger self-evaluations of at least valence and arousal. In AD, driver behavior is mainly relevant for end-to-end driving models, not for learning which situations and vehicle behaviors influence passenger UX. A driver always has control over the vehicle's actions and therefore over one's own UX, while a passenger has little to no control over the vehicle's actions and therefore cannot influence one's own UX. This presents a key research gap to bridge AD data and affective computing.

### 2.3 SoTA Methods for multi-modal HMI datasets

Datasets on HMI especially with both self-assessment data, multi-modal user sensor data, and machine data are scarce due to several factors. Mainly, real-time self-assessment methods are missing and real-user data with e.g. physiological signals are valued as sensitive data and are strongly protected by data protection policies. Additionally, most datasets describe approaches to collect driver fatigue and physiological signals, but do not incorporate self-assessment methods during a ride, since drivers are not treated as passengers (e.g. UPCT-dataset [34], AffectiveROAD [10], PPB-Emo [24]). Finally, the collected datasets focus on driver behavior and simulation (e.g. Tao et al. [41], UPCT [34]) and are not conducted in real-world AVs. Therefore, a key research gap addressed in our work is not in delivering a dataset, but in providing the scientific community with a study protocol to evaluate their own autonomous or robotic systems based on user-centric acceptance data.

# 3 METHODOLOGY

Our research goal is to create a study protocol supported by real-world field studies and sensor suggestions, including code for each sensor data collection. Based on findings of each study, a final study setup is created.

## 3.1 Pre- and Post-Questionnaire

A widely used and effective method of UX assessment is questionnaires [23]. Therefore, most of our field-studies include an overall evaluation of the UX based on a combination of different questionnaire-models. To assess acceptance in a questionnaire we identified the following models in literature relevant for our use case: TAM [4], UTAUT [42], CTAM [29] and AV Acceptance Model (AVAM) [14]. We only use components of the AVAM in our questionnaire since each model mentioned above is a modification of the preceding models.

The intention to use AVs or new technologies depends on one's technological openness [8]. To address the criticism in literature on the lack of transparency regarding the fundamental attitudes of participants towards AVs and new technologies [22, 27], we conducted a preliminary survey (*pre-questionnaire*). Before the interaction with the vehicles, participants provided their agreement to 17 statements, which were divided into three sections based on studies found in literature [15, 17, 25, 33]. The first section inquired about attitudes toward AD, the second about attitudes toward AI and new technologies, and the third about general trust in machines. These three sections were considered as a construct for general technology acceptance(TA). Based on their answers, we classify participants into four distinct groups (Rejectors, Conservatives, Pragmatists, and Enthusiasts) following [15] and [45]. This enables us to determine whether the results hold validity beyond the group of individuals with a positive attitude toward AVs and new technologies, which are often overrepresented in many studies.

After the interaction participants fill out a post-interaction questionnaire (*post-questionnaire*). The post-questionnaire contained questions on the following constructs: UX, Perceived Safety, Trust, Ease of use, Transparency, Sense of control, Enjoyment-fun-and-interest, overall rating of the complete interaction, Net-Promoter Score (NPS), willingness to pay for a 10-minute interaction, and an open question for further comments. *The questionnaire can be found here:* `https://gitlab.kit.edu/kit/aifb/ATKS/public/AutoSMiLeS/questionnaire`.

**Analysis:** We calculate the mean TA across the *pre-questionnaire* to obtain an overall assessment of participants' TA construct, following [15, 45]. Each question is answered on a Likert scale from 1 to 5. A mean $\geq$4.5 assigns the participant to the Enthusiasts, $\geq$3.25 – to the Pragmatists, $\geq$2.75 – to the Conservatives, and $\leq$2.75 – to the Rejecters group.

For the *post-questionnaire*, the number of participants required to achieve statistical significance for UX was analyzed by [5]. The median number of participants in 487 studies using UEQ, AttraktDiff and meCue was 20, with 12 in

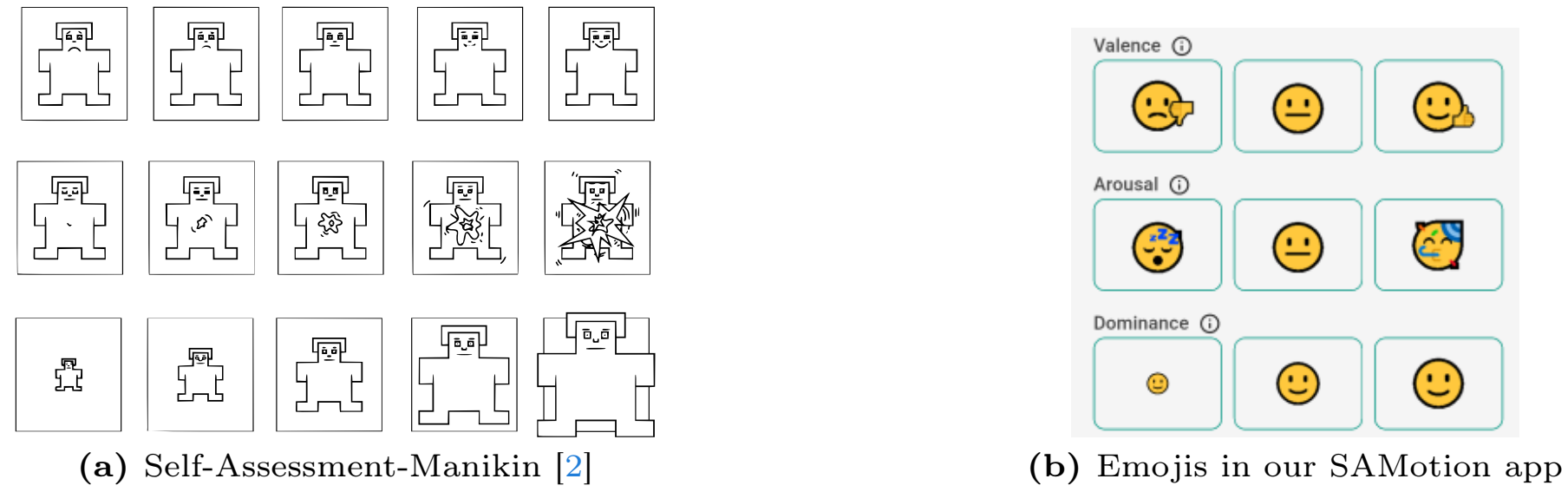

**(a)** Self-Assessment-Manikin [2]

**(b)** Emojis in our SAMotion app

**Fig. 1:** Visual comparison of (a) SAM as defined in [2] and our (b) SAMotion app.

the lower quartile and 30 in the upper quartile. The authors of the UEQ and UEQ-S also provide a "Sample Size Estimator"[4].

For validation of the questions and therefore the questionnaire we calculated Cronbach's Alpha for each construct mentioned above. We used the Mann-Whitney U test to measure the statistically significant difference between the distributions of two independent samples of a control group and a validation group. In a second step, we calculated the effect size according to Cohen for the statistically significant constructs. According to Cohen [3], a value of $|r| \geq 0.1$ indicates a weak effect, $|r| \geq 0.3$ indicates a medium effect, and $|r| \geq 0.5$ indicates a strong effect of the difference between the groups.

### 3.2 SAMotion - App for Self-Assessment during Interaction

Most user studies provide questionnaires with standard assessment methods such as SAM (Self-Assessment-Manikins) [2] after interactions with machines or facial-expression-based affect evaluations with no self-evaluation [46]. However, self-assessment of one's own affect in hindsight does not provide the exact time of the feeling and is also prone to decreasing accessibility of detailed memory over time. In addition, similar facial-expressions do not always convey consistent affect inter- and intraculturally [1]. A key research gap missing in literature on user studies in HMI is a real-time self-assessment method. Therefore, we developed a smartphone app that can communicate directly with the machine (over ROS messages - Robot Operating System) and provides a user interface for users to evaluate their affective states. Touch events are recorded directly, sent to the machine via websockets over WLAN and rosbrigde server[5], and saved there to evaluate machine and user data together. This allows for machines, e.g. AVs, to incorporate not only machine data into planning and prediction, but also human affect. This development is an important step towards the successful integration of machines into human environments, which increases trust, understanding, and collaboration.

[4] https://www.ueq-online.org/Material/Data_Analysis_Tools.zip
[5] https://wiki.ros.org/rosbridge_server

**Table 1:** Overview of the conducted studies.

| # | Interaction | Location | Sensor Setup | Number of Participants | User data | | | | Assessment |
|---|---|---|---|---|---|---|---|---|---|
| | | | | | Camera | Audio | IMU | Physiological | |
| 1 | Shuttle | 30 km/h, real traffic | | 36 | | | | | Questionnaire |
| 2 | Video | Indoor | Apple Watch, Muse | 11 | | | | EEG, heart rate | Questionnaire, SAMotion |
| 3 | Shuttle | 30km/h, campus | Zed 2i, Muse | 6 | ✓ | | | EEG | Questionnaire, SAMotion |
| 4 | Car | 50 km/h, real traffic | Apple Watch, Muse, breath band | 10 | | | | EEG, heart rate, breath | Questionnaire |
| 5 | Car | 50 km/h, real traffic | Samsung Galaxy S23 | 16 | ✓ | ✓ | | | |
| 6 | Bicycle | 50 km/h, real traffic | Samsung Galaxy S23 | 18 | ✓ | ✓ | | | |
| 7 | Shuttle | 30 km/h, real traffic | Samsung Galaxy S9 | 14 | ✓ | ✓ | ✓ | | Questionnaire, SAMotion |
| 8 | Car | 50 km/h, real traffic | Samsung Galaxy S9, iPhone 14, breath band | 32 | ✓ | ✓ | ✓ | Breath | Questionnaire, SAMotion |

In SAM, the three dimensions of valence, arousal, and dominance are each represented with five manikin images (see Figure 1a). However, choosing between five manikins for each VAD-dimension during an interaction takes up a lot of time. Therefore, we reduced the number to three manikins and adapted them to emojis (see Figure 1b). During an interaction, each situation can be assessed based on one's own affect. The code for SAMotion Android and iOS can be found here: https://gitlab.kit.edu/kit/aifb/ATKS/public/AutoSMiLeS/samotion.

# 4 EXPERIMENTS

Our work comprises eight studies with 144 participants in sum (see Table 1). First, we describe the AVs used (see Figure 2). Then each study is described by motivation, research goals, data collection setup, results, and, finally, insights to be adapted for the succeeding studies. Figure 3 shows dependencies of learnings and sensors per study.

**Autonomous Shuttle:** The autonomous shuttle is an EasyMile vehicle equipped with our own sensors[6] and autonomous software [28] for SAE Level 4 autonomy and sensory systems, including five LiDAR sensors and cameras (see Figure 2b). It has 3 seats in one direction and 3 in the other. In all the studies conducted, passengers sat on only one of the three seats facing the driving direction. In Germany where the studies were conducted, a mandatory safety operator is required by law. The software takes over all driving functions.

**Autonomous Car:** The used car was CoCar NextGen [12], a commercial Audi A6 Avant Plug-in-Hybrid, which was modified with additional sensors, actuators for braking, steering, and acceleration, and the same autonomous software [28] as in the autonomous shuttles presented above (see Figure 2a). A legally mandatory safety operator was able to take over driving functionalities at any moment at the press of a button. A steering wheel and pedals for breaks and gas were still in the car.

[6] https://www.fzi.de/en/research/research-infrastructure/fzi-shuttles/

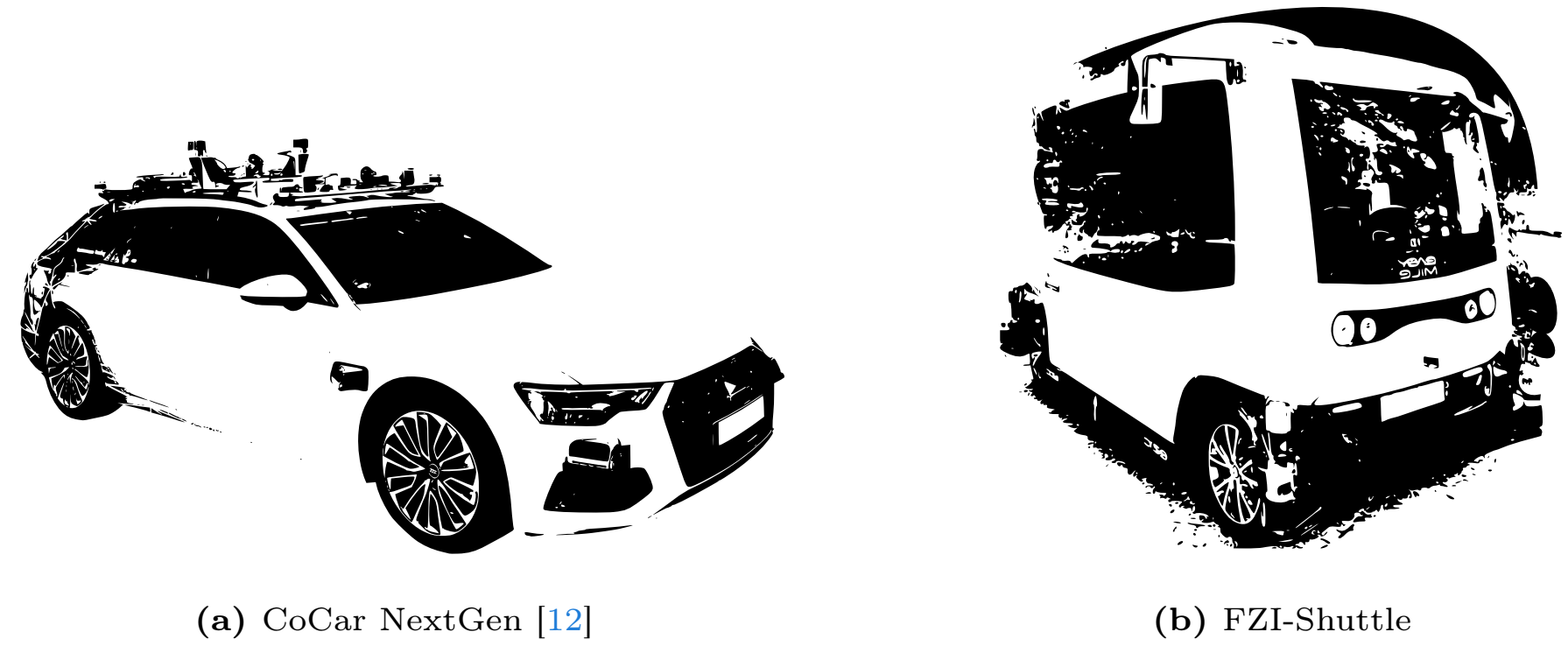

**(a)** CoCar NextGen [12] **(b)** FZI-Shuttle

**Fig. 2:** Schematic representation of the AVs used in our studies.

### Study 1: Autonomous shuttle with questionnaire

The **goal** of the study was to establish whether a questionnaire is sufficient for UX evaluation and to validate the developed questionnaire for further studies.

**Setup:** There were 18 participants in a test group (with driving data visualization) and 18 participants in a validation group (without visualization). All experienced a ride in the autonomous shuttle of approximately 8 minutes and the same route in a 30km/h zone in real traffic, and all completed the pre- and post-questionnaire.

**Results:** 36 passengers participated in the study and all answered the questionnaires. We used the Mann-Whitney U test to assess whether there is a statistically significant difference between the distributions of the test and validation groups. By comparing the mean rank sums, the direction of the difference can also be interpreted. There was a statistically significant difference in ease of use ($Z=-3.040$, $p=0.005$), perceived enjoyment, fun, and interest ($Z=-3.040$, $p=0.005$), and perceived sense of control ($Z=-3.302$, $p<0.001$) between participants in the test and validation groups. Comparison of mean rank sums shows that participants in the test group gave a significantly better rating for all three items, therefore showing that an additional visualization in the AV improves the understanding of the actions of the AV as well as the ease-of-use, including the enjoyment, fun, and interest, and the perceived sense of control. For all other scales, no significant differences were observed between the test and validation groups using the Mann-Whitney U test.

**Insight:** This study showed that a visualization of the AV's actions creates a statistically significantly higher UX. These results demonstrate that a questionnaire can be used to evaluate an entire HMI experience. Additionally, the evaluation validated our developed questionnaire. However, no direct evaluation of different scenarios during the ride was possible. This made automatic evaluations of specific driving functions impossible.

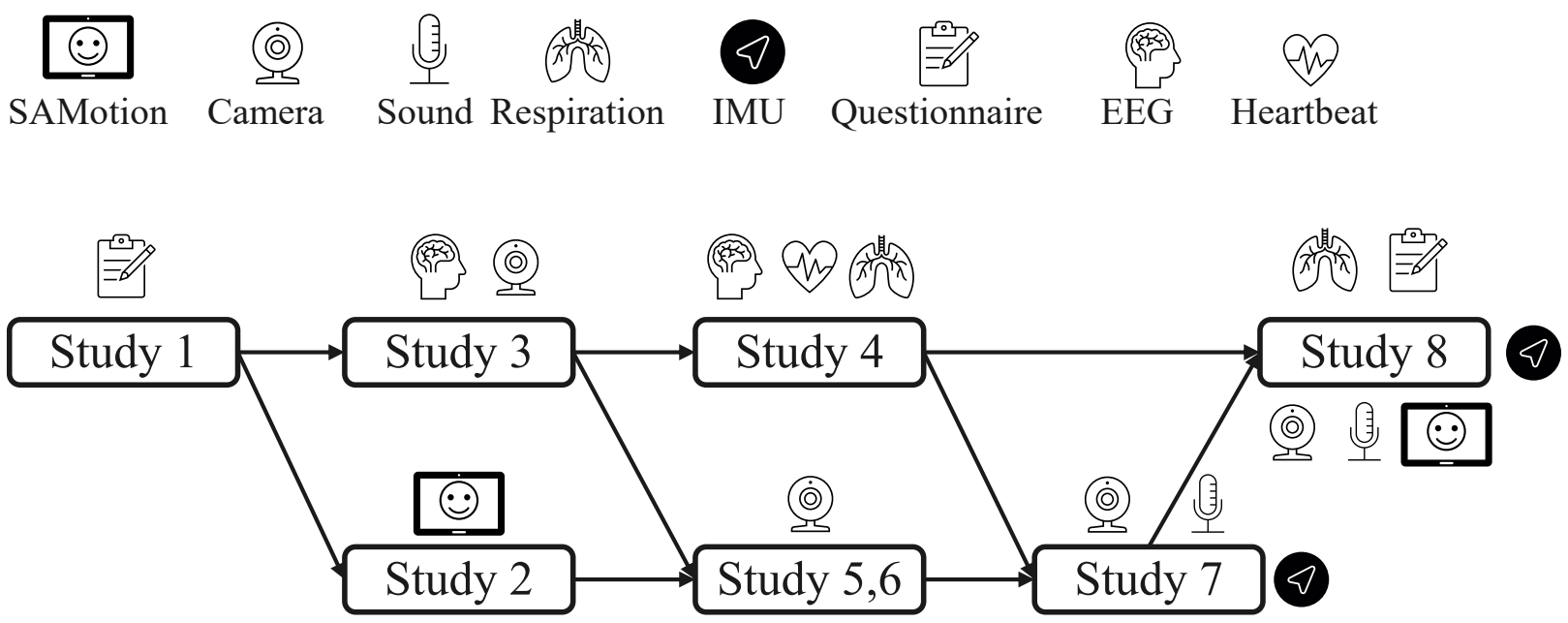

**Fig. 3:** Study dependencies based on results and insights gained from preceding studies. The mainly analyzed sensors per study are shown.

### Study 2: Video validation for SAMotion

Based on learnings from the first study, we developed the SAMotion app (see Section 3.2) for iOS and Android, where each participant can evaluate ones own affect in real-time during an interaction with a machine. This self assessment is then sent to the machine (i.e. AV) itself. The **goal** of the second study was to evaluate the app based on its validity in capturing human affect.

**Setup:** We set up a video study according to DREAMER [20], where participants watched videos based on [9] which can induce affective and emotional states. After each video, the participants rated their affective states using SAM or SAMotion. Both ratings per movie-clip were compared based on Spearman's rank correlation coefficient ($r_s$) and its corresponding p-value (p<0.05 statistically significant value). Also, EEG-data was collected for further analysis of objective measurement methods in study 5.

**Results:** 11 participants watched videos, in order to validate the reliability of our SAMotion app for a participant's self-assessment in comparison to the widely used method of SAM. Six participants evaluated movie clips based on SAM and five based on SAMotion. Our results show a strong correlation between both assessment methods for the dimensions valence ($r_s$ = 0.88; p = 0.00155), arousal ($r_s$ = 0.71, p = 0.0334) and dominance ($r_s$ =0.9; p = 0.00103) and therefore a validation of using SAMotion instead of SAM.

**Insight**: The results demonstrate the reliability of SAMotion for self-evaluation in HMIs. It closes the research gap of evaluating one's own affect during an interaction with a machine and sending this information to the machine itself.

### Study 3: Autonomous shuttle with a camera and EEG

Due to the subjectivity of the self-assessment, additional objective measurements, such as physiological signals, were evaluated in study 3 for correlations to UX. These are an EEG and a camera. The used camera was a depth camera (ZED 2i), with a framerate of 30Hz. We used a commercial EEG measurement headband Muse S[7] with a framerate of 256 Hz. It includes only four EEG-Channels: TP9, AF7, AF8, and TP10. The **goal** of the study was to create a small dataset from low-cost consumer EEG devices, a camera, and a machine simultaneously and train simple models based on machine data with output of passengers' self-assessment data and attention.

**Setup:** Participants sat in the same autonomous shuttle, but this time the route was placed in a controlled campus environment where little traffic and few pedestrians frequented. Participants put on the headband and were recorded by the camera. During the ride, participants were asked to use the SAMotion app and evaluate situations based on the VAD dimensions. For each participant, the same route was used, and one single cyclist and one single car were used for two overtake interactions with the AV. In no instance was there a chance of collision. However, the subtle but stronger braking of the shuttle was used for passengers attention shift.

**Results:** Our results show, that not using a bandpass filter on EEG improved accuracy results by 10% of predicting whether a person was looking at a tablet or not. We compared a Random Forest Classifier (RFC), gradient boosting (GB), Support Vector Machine, an ensemble learning approach of all three (RFC, GB and SVM) and a Gated Recurrent Unit (GRU). Best results on accuracy were received by the GRU with 84% on the validation dataset.

**Insight**: Model training of attention on a tablet or attention on the road with EEG is possible, with a very limited amount of data. Additionally, using no bandpass filter (raw data) instead of removing artifacts such as eyeblinks from the EEG data resulted in better model performance. Possibly indicating artifacts as important prediction factors or a certain noise is created during head-movements on the sensors and this noise is learned by the model. However, using the Muse-S headband for EEG-measurement proved challenging due to sampling rate variations and timestamp inconsistencies.

### Study 4: Autonomous car with EEG, heartbeat, and breathing

Results of studies 2 and 3 indicated a minor correlation of EEG data, attention on the smartphone, and the self-assessment VAD-dimensions. However, more training data were needed and a comparison to models with larger numbers of EEG-channels would be beneficial, to indicate the real potential of EEG in affect-inference. The **goal** of this study was to evaluate prediction of affect based on autonomous car data and physiological signals. Therefore pretrained models of larger EEG-datasets including breathing and heartbeat were created in our previous work [16] and fine tuned on self collected four-channel EEG-data, heartbeat and breathing belt.

**Setup:** The study was conducted in real-world traffic in a high traffic 50 km/h zone with the autonomous car. The used breathing belt was the Go Direct Respiration Belt by Vernier[8]. The data was transmitted over Bluetooth. Code for respiration belt data collection can be found here: https://gitlab.kit.edu/kit/aifb/ATKS/public/AutoSMiLeS/respiration. In our setup, the data was sent to a laptop which then transmitted it to the car computer, since the participant was sitting in the front passenger seat and the computer was located in the trunk. The Bluetooth connection would have been unstable due to several obstacles in between, such as car seats and different sensors.

[7] https://choosemuse.com/
[8] https://www.vernier.com/product/go-direct-respiration-belt/

Participants also put on the EEG-headband (Muse S) and the Apple Watch for heart-rate monitoring. Since we did not want to obstruct the data with head movements which were not due to the car movements or environment, we opted not to include the SAMotion app so as not to induce head movements towards the smartphone or tablet. The safety operator was sitting behind the wheel.

**Results:** Minor correlations of EEG and machine data were observed. However, an evaluation in our previous work [16] of a larger DEAP EEG dataset [21] showed significantly higher results with setups that contained at least 12 electrodes. Because more complex setups do not scale well for user studies, EEG was not used for affect inference in subsequent studies.

Heartbeat evaluations did not show any significant correlations towards machine data and affect, and were therefore also not considered for the final study setup in study 8.

However, breathing rate evaluations showed significant correlations between participants' highest 10 breathing rates at similar locations on the driving route. Analysis of driving scenes revealed interesting events at these breathing-peak hotspots. However, some hotspots contained no indications of interest. Since we did not have recordings of the camera or the passengers' voices, some breathing events could have been similar to situations in which participants communicated with the safety operator. Therefore, voice-recordings were also considered for the final study 8.

**Insight**: Affect inference over EEG is possible, but was dismissed due to inconsistencies in headband data and the difficulty of applying more complex EEG hardware setups, especially with extruding electrodes on the head during hard-break events. Affect inference over heartbeat showed discordant results and was also dismissed for the final study. Correlations between breathing rate and AV actions were identified through simple procedures but require additional information from voice recordings for the final study.

### Studies 5 and 6: Video for head-pose recognition

Study 3 showed promising results for modeling whether passengers attention was on a smartphone or on the AV and its surroundings. The next experiments aimed to evaluate precise head movements to understand where a person was looking and what the person saw. An iPhone 14 was used to collect video data from head movements. The **goal** of study 5 was to create a small dataset to evaluate head-pose models based on head-movements during a ride with an autonomous car.

**Setup 5:** The same route, car, and autonomous mode was used from study 4. Participants sat in the front passenger seat and a safety-operator sat behind the wheel.

Similarly to study 5, the **goal** of study 6 was to validate head-pose models on real video and IMU data. However, here data was collected during a ride with a bike in order to evaluate whether a generalization of head-pose models is possible from the car-scenario to the bike-scenario.

**Setup 6:** Participants rode a bike with an iPhone 14 positioned on the handlebar of the bike. The route each participant took was the last part of the autonomous car route in study 5. Each participant traveled the same route for approximately 20 minutes.

**Results:** For studies 5 and 6 the goal was to fine-tune existing head-pose models on camera data recorded inside the AV and also on camera data recorded on a bicycle in order to train a more generalized model. The purpose of a general model was to record where a persons attention was during the study to connect a persons affect with car actions. HopeNet [35], 6DRepNet360 [13] and WHENet [44] were fine-tuned and compared on car, bike and recordings of both.

**Insight**: Highest results were achieved by HopeNet with a fine-tuning of data from both car and bike data (accuracy 97.31%) and tested on car data. Therefore, the camera is also used to predict the attention for the final study setup.

### Study 7: Autonomous shuttle with SAMotion and questionnaire

Study 1 showed a correlation between UX and the addition of a visualization of the shuttles' data. Studies 5 and 6 showed promising results for smartphone-based head-pose recognition. Therefore, the **goal** of this study was to evaluate UX during an autonomous ride based on two different visualizations using camera, SAMotion and questionnaire. Additionally, we tested whether one camera on a handheld tablet was sufficient for head-pose recognition or whether a mounted camera similar to that in study 5 is needed.

**Setup:** Participants were chauffeured along a similar route as in study 1. They used the same autonomous shuttle, but in addition to a questionnaire, participants were given a Samsung Galaxy S9 tablet with the SAMotion app to assess situations during a ride based on their affect. Camera and audio were also recorded based on results from study 4.

**Results:** Two visualizations were evaluated based on questionnaire (Mann-Whitney U-test) and self-assessment during the ride. A significant difference was found in UX hedonistic quality between participants in the group 1 and 2 ($Z=-2.239$, $p=0.025$). The Z-value indicates that group 2 rated higher, as expected from the mean value comparison. Group 2 also scored higher on the understandability of interaction ($Z=-2.946$, $p=0.003$), overall experience of the ride ($Z=-3.064$, $p=0.002$), using AVs in public traffic ($Z=-2.475$, $p=0.013$), and the Net-Promoter Score ($Z=-2.239$, $p=0.034$). All scales showed large effect sizes, with all r-values above 0.5.

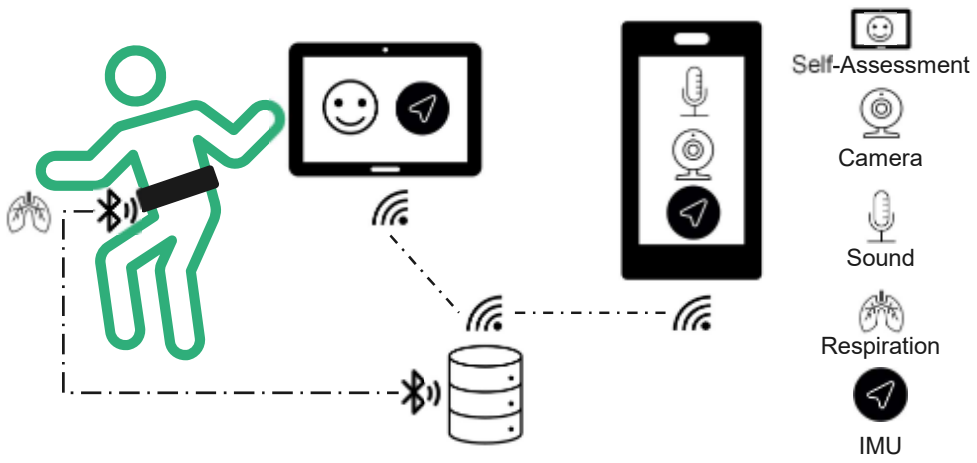

**Fig. 4:** Study 8 setup: user data sensors (camera, sound and IMU over iPhone 14, Go Direct respiration belt), SAMotion app on a Samsung Galaxy S9 tablet. Data transmission over wlan and bluetooth to the onboard autonomous software server (bottom).

**Insight**: All in all, visualization 1 (added autonomous-mode indicator and map visualization) showed a significantly lower UX than visualization 2 (only speed indicator and LiDAR-visualization), contrary to expectations. This phenomenon could be explained by social desirability bias of subjects in group 2 or higher occurrences of technical difficulties in group 1. Further analysis on the two designs need to be evaluated. Additionally, holding a tablet and directly recording video data of the face from the upward angle is not sufficient for facial affect inference and head-pose detection. Therefore a further mounted camera is needed and added to the final study setup.

### Study 8: Final evaluation in autonomous car

Self-assessment evaluations show interesting peaks. First, participants mostly rated events with either negative or neutral valence and near to no positive valence (see Figure 5a). Second, aggregated SAMotion rating peaks over all participants (see Figure 5b) show similar peaks at similar points along the route as the aggregated respiration peaks (see Figure 6b). The SAMotion hotspots occur shortly after the respiration hotspots, indicating a delay in self-evaluation compared to reactions of the autonomous nervous system in the form of increased respiration.

Learnings from previous studies showed promising results for camera, audio, tablet IMU, and respiration belt. Therefore, the final study setup consists of these user data sensors, including machine data and assessment through a questionnaire, and our SAMotion app. An overview of the final setup can be seen in Figure 4. The **goal** of this study was to collect a dataset to evaluate the autonomous software based on UX.

**Setup:** The autonomous car from studies 4 and 5 was used in real-world traffic in a 50 km/h city zone. Study 8 contained the final study setup with sensors based on all the learnings from the previous studies. We used the user data sensors of a mounted smartphone (iPhone 14) attached to a phone holder on

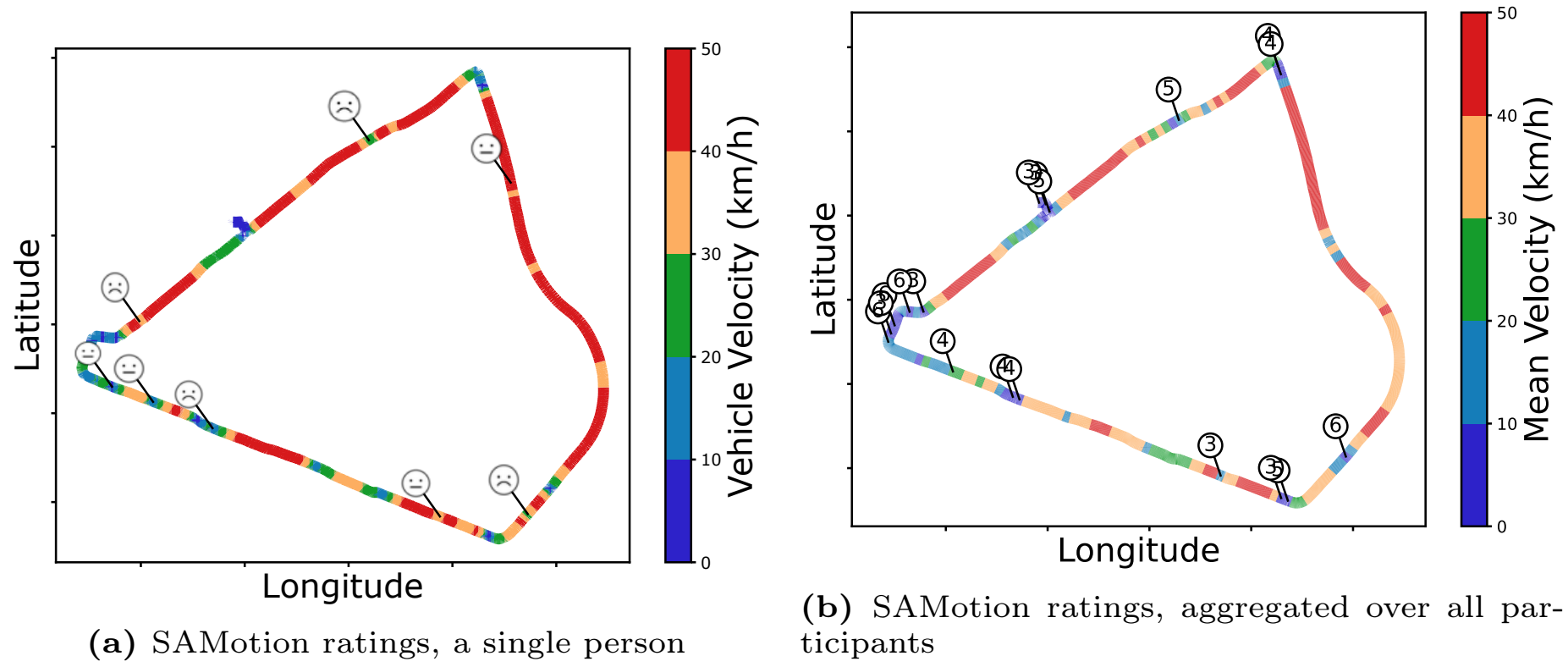

**(a)** SAMotion ratings, a single person

**(b)** SAMotion ratings, aggregated over all participants the front windshield. It recorded sound, camera (participant's head), and IMU over settings of the SAmotion app. A handheld tablet (Samsung Galaxy S9) was used to record the self-assessment with the SAMotion app of each participant during the autonomous ride including IMU data. In addition, the respiration belt from study 4 was used. All devices communicate with the onboard server that contains the autonomous software. The communication with the autonomous software was done via WLAN. The participant sat in the front passenger seat.

**Fig. 5:** SAMotion evaluations along the route for Study 8: **(a)** each emoji represents negative, neutral, or positive valence indicating an interesting event occurrence, while the size of the emoji represents the degree of arousal. **(b)** SAMotion rating occurrences of all participants aggregated into hotspots with a 25 m distancing, combined with color coding for the mean vehicle velocity over all participants in Study 8. The numbers indicate how many participants rated an event at the respective location.

**Results:** Route-evaluations show interesting respiration peaks. For each participant, the 10 highest respiration frequencies along the route were assessed. An exemplary evaluation for one participant can be seen in Figure 6a. The 10 respiration peaks are numbered according to their occurrence in time. Then, all respiration peaks of all participants are aggregated along 25 m distancing blocks. The respiration hotspots along the route can be seen in Figure 6b) aggregated among all participants. It should be noted that there are nearly no hotspots in the yellow and red zones, where the AV drove more than 30 km/h. This may be due to the fact that respiration peaks are aggregated along a distance of 25 m, which is less than 3 seconds in duration when driving above 30 km/h. However, those hotspots that do exist in the above 30 km/h zones are noticeably more interesting. For example, in Figure 6b in the lower middle, there is a hotspot in a yellow zone where 6 participants exhibited one of their 10 respiration peaks along the route. Looking closer into the vehicle data shows a lane change in this area where the vehicle is closer than usual to the boarder-stone. All other hotspots occur in junctions with or without traffic lights.

Voice evaluations show that the transcription and speaker-diarization models have improved over the years. Speaker-diarization is needed to identify whether

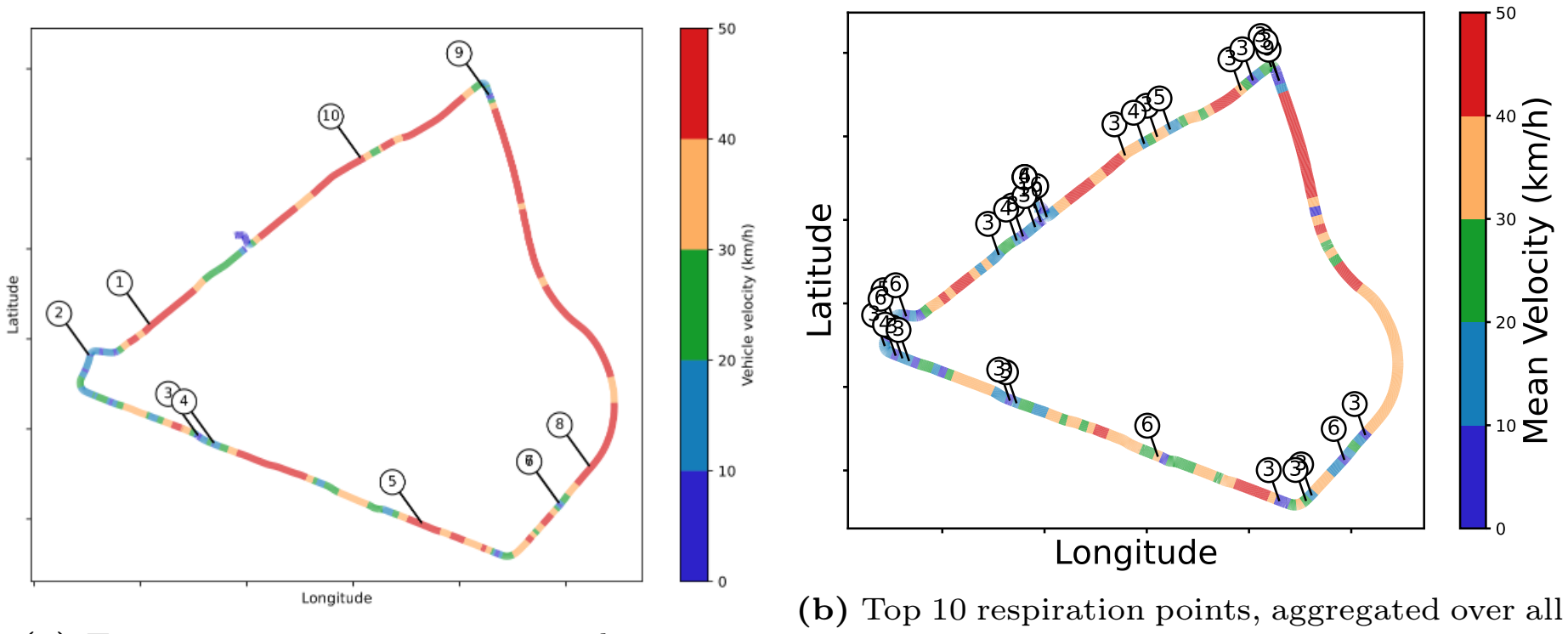

**(a)** Top 10 respiration points, a single person

**(b)** Top 10 respiration points, aggregated over all participants the passenger was speaking or the safety operator in the autonomous vehicle. We tested two methods for transcription and speaker-diarization. The first approach included the multilingual whisper-large model [32] (from 2022) for transcription, while the resulting speaker embeddings were used for speaker-diarization through a k-means clustering. However, this method resulted in more than 50% of the passengers having hallucinated transcriptions and, therefore, hallucinated diarization. For the second approach, we used the vibevoice-asr model [31] (from 2026) that included speaker-diarization in its output. There were some repeated words in the output, but no hallucinations. A synchronization between smartphone and car data shows a mean delay of 200 ms, which is due to wlan latency in the car. In general, the GPS-time from car and smartphone should be nearly identical and the delay of 200 ms needs only be accounted for, when models of transcription, diarization and facial evaluations are run on the on-board server online instead of evaluated afterwards.

**Fig. 6:** Respiration evaluations along the route for Study 8: **(a)** each number along the route represents one of the ten highest respiration rates for a participant, indicating an interesting event occurrence. **(b)** respiration-rate peaks (top ten per participant) aggregated into hotspots with a 25 m distancing, combined with color coding for the mean vehicle velocity over all participants in Study 8. The numbers indicate how many participants had one of their top ten respiration peaks at the respective location.

## 5 CONCLUSION

Our work outlines goals and challenges for real-world studies and provides insights for the scientific community to improve robotics through user-centric methods. These methods are shown and tested on studies with AVs in real-world traffic but are applicable to the whole field of robotics. We invite the scientific community to incorporate humans in the development process and not only focus on the technical implementation, improving acceptance and UX of robots in everyday lives.

## Acknowledgements

This work was supported by funding from the Topic Knowledge for Action of the Helmholtz Association (HGF).